\documentclass[aps,prd,twocolumn,showpacs,preprintnumbers]{revtex4}
\usepackage{graphicx,bm}
\usepackage{dcolumn}    
\usepackage{color}

\definecolor{red}{rgb}{1,0,0}
\definecolor{blue}{rgb}{0,0,1}
\definecolor{green}{rgb}{0,1,0}

\begin{document}

\title{Black hole production in tachyonic preheating}
\author{Teruaki Suyama$^a$, Takahiro Tanaka$^a$, 
Bruce Bassett$^b$ and Hideaki Kudoh$^c$}
\affiliation{$^a$Department of Physics, Kyoto University, Kyoto 606-8502, Japan,\\
$^b$Institute of Cosmology and Gravitation, University of Portsmouth, Mercantile House, Portsmouth PO1 2EG, UK and SAAO/UCT, Observatory, Cape Town, South Africa\\
$^c$Department of Physics, The University of Tokyo, Bunkyo-ku, 113-0033, Japan}


\begin{abstract}
We present fully non-linear 
simulations of a self-interacting scalar field in the early
universe undergoing tachyonic preheating. 
We find that density perturbations on sub-horizon
scales which are amplified by tachyonic instability
maintain long range correlations even during the succeeding
parametric resonance, in contrast to the standard models 
of preheating dominated by parametric resonance. As a result the final 
spectrum exhibits memory and is not universal in shape.  
We find that throughout the subsequent era of
parametric resonance the equation of state of the 
universe is almost dust-like, hence the Jeans wavelength
is much smaller than the horizon scale. 
If our 2D simulations are accurate reflections of the situation in 
3D, then there are wide regions of parameter 
space ruled out by over-production of black holes. It is likely however 
that realistic parameter values, consistent with COBE/WMAP normalisation, 
are safetly outside this black hole over-production region.
\end{abstract}


\preprint{KUNS-2007}
\preprint{UTAP-527}  

\maketitle

\section{Introduction}

Over the past decade, the theory of reheating
after inflation has been developed and it was
recognized that reheating can be 
accompanied by an era of preheating, 
where fluctuations of scalar fields are 
greatly amplified either by parametric
resonance \cite{Dolgov:1990,Traschen:1990sw,Fugisaki:1996,Kofman:1994rk,Kofman:1997yn}
or by tachyonic instability \cite{Green:1997,Felder:2000hj,Felder:2001kt}. In preheating, energy transfer from the inflaton field to another field occurs in bursts where non-perturbative effects are important for the evolution of scalar fields.

A common feature among the various preheating models is that preheating leaves an imprint on the subsequent evolution of the universe, by amplifying the density perturbation especially
on sub-horizon scales at that time by many orders of magnitude. 
If the resultant density perturbations at horizon crossing becomes of order unity, black holes can be formed by gravitational collapse \cite{Carr:1974nx}.
Since the energy density of produced BHs increases proportional to the scale factor relative to the total energy density of the universe in the subsequent radiation dominated era,
BHs will dominate density of the universe at late time even if a tiny fraction of energy is converted into BHs at preheating. There are strong constraints on the abundance of 
BHs for a wide range of mass scales \cite{kohri} and they can be a powerful means for constraining inflationary models \cite{Yokoyama:1999xi}.

In the past, it was pointed out by several authors
that BHs can be over-produced during preheating 
for two-field models \cite{agreen, bassett,finelli}.
In these models, parametric resonance causes the rapid amplification of field fluctuations.
Then, non-perturbative effects such as rescattering may be crucial for calculating the abundance of produced BHs. 
Indeed, it was shown via lattice simulations that BHs are unlikely to be over-produced during preheating \cite{Suyama:2004mz}. Accurate simulations are required since a small error in the estimate of backreaction time which ends the resonance causes a large error in the amplitude of density perturbation at the end of preheating because field fluctuations grow exponentially during preheating. 
 
In this paper, we use lattice simulations to examine the evolution of density perturbations during tachyonic preheating and the succeeding parametric resonance, mainly focusing on the production of black holes.
To evaluate the amplitude of density perturbations taking the non-linear interactions into account, we performed two and three-dimensional lattice simulations using the  
C++ code LATTICEEASY developed by Felder and Tkachev \cite{latticeeasy}. 

The equations of motion of the minimally coupled scalar field that we solve are
\begin{eqnarray}
&& {\ddot \phi}+3 H {\dot \phi} -a^{-2}\triangle {\phi} +
 V'(\phi)=0, \\
&& H^2 = \frac{8\pi}{3 M^2_{pl}} \langle \rho \rangle,
\end{eqnarray}
where $ \rho $,  $ a $ and $H$ are, respectively, the total energy density, 
the scale factor and the Hubble parameter, and 
$ \langle \cdots \rangle $ denotes spatial average. 
To estimate the variance of density perturbations at the horizon scale ($k=aH$), we simply use the power spectrum ${\cal P}_{\delta} (aH)$. 
Here ${\cal P}_{\delta} (k)$ is defined by
\begin{equation}
\langle {\delta}_k {\delta}_{-k'} \rangle = \frac{2 {\pi}^2}{k^3} {\cal
P}_{\delta} (k) \delta ( \vec{k}-\vec{k'}), \label{pbh5} 
\end{equation}
and $\delta_k$ is the Fourier component of the density contrast $ \delta \equiv \delta \rho/\langle \rho \rangle $.

\section{Tachyonic preheating}

It was noted in the lattice simulations of \cite{Felder:2000hj,Felder:2001kt} that the tachyonic instability in the context of spontaneous symmetry breaking can be the most efficient process of energy transfer from the potential energy of scalar field into the energy of its inhomogeneous modes.
If $ V''(\phi) $ is negative, fluctuations with comoving momenta $k < \sqrt{|V''|} $ grow exponentially and can  often become quite large. As a result of backreaction, the spatial average of the field experiences only a few oscillations around the minimum of the potential or may not even oscillate at all. 

One important characteristic in tachyonic instability is that the longer wavelength modes have the largest growth rate as long as the metric perturbations can be neglected, which will be a good approximation for the dynamics inside the horizon. On the other hand, it is well known that curvature perturbations on a constant energy density hypersurface remain constant on super-horizon scales if the evolution of the energy density is described solely by a single equation of state.  Thus no tachyonic instability occurs far outside the horizon in single scalar field models \cite{Bassett:1999mt}, though we expect that perturbations on sub-horizon scales are amplified by tachyonic instability, which may still lead to over-production of BHs. 
Therefore it is interesting to see how large the density perturbations can be amplified become through the tachyonic instability on sub-horizon scales where one can neglect metric perturbations. 

Here we consider Coleman-Weinberg potential \cite{Coleman:1973jx} 
\begin{equation}
V(\phi) = \frac{1}{4}\lambda {\phi}^4 \left( \log \frac{\phi}{v} -\frac{1}{4} \right) +\frac{1}{16} \lambda v^4, \label{tachyon1}
\end{equation}
which was used in the original model of new inflation 
\cite{Linde:1981mu,Albrecht:1982wi}.
Recently it was found in \cite{Desroche:2005yt}
that preheating in this model is so efficient 
that the perturbative description of 
the reheating process \cite{Dolgov:1982th,Abbott:1982hn}
does not apply.

To see for which parameters of the potential BHs
are most likely to be produced, 
let us consider the growth of the 
density contrast $\delta$
in the linear approximation.
Approximately 
when the background $ \phi $ reaches the value
$ {\phi}_o = v (v/M_{pl}) $, the
negative mass squared of the inflaton field, $-V''$, 
exceeds the squared Hubble parameter, $H^2\approx \lambda v^4/M_{pl}^2$. 
After this epoch, tachyonic instability sets in. 
Therefore, when $ \phi = {\phi}_o $, one can still set 
the initial amplitude $ \delta \phi$ to  
$\sim H\sim \sqrt{\lambda}v^2/M_{pl}$ 
at around the horizon scale ($ k/a \sim H $).  
On large scales the growing mode of ${\delta \phi}$ 
grows in proportion to ${\dot \phi}$.
Neglecting the expansion of the universe,
which will be valid when the mass scale exceeds 
the Hubble scale ($\phi > {\phi}_o $), one can say that 
${\dot \phi}$ grows in proportion to ${\phi}^2$ from energy conservation \cite{Desroche:2005yt}. 
Therefore at a small $ k $ we have $ {\delta \phi} \propto {\phi}^2 $.

Using the relations mentioned above, one can deduce 
$\delta\phi/\phi \sim \sqrt{\lambda}M_{pl}\phi/v^2$. 
In the parameter region $\sqrt{\lambda}<(v/M_{pl})$, $\delta\phi/\phi$ stays less than unity. 
Therefore fluctuations do not become non-linear. 
In this case the amplitude of density perturbations also will not reach the $O(1)$ values required for BH production. 
On the other hand, in the case where
$\sqrt{\lambda}>(v/M_{pl})$, fluctuations of the inflaton field evolve into non-linear regime.

If we assume that primordial density perturbations observed, e.g. in the anisotropy of the cosmic microwave background, are produced from the vacuum fluctuations of 
$ \phi $, $ \lambda $ is fixed to about $ 10^{-12} $ from COBE (COsmic Background Explorer) normalization.
If we adopt this value, the most interesting parameter for $ v $ is $ 10^{-6} M_{pl} $ or less from the above argument.
But in this case, there is a large hierarchy (six orders of magnitude) between the minimal wave number for tachyonic instability to occur, which is determined by the horizon scale 
($ \sim \sqrt{\lambda} v (v/M_{pl}) $), and the one for parametric resonance to occur 
($ \sim \sqrt{\lambda} v $). 
Therefore we cannot incorporate these two scales simultaneously into our numerical simulations for the parameter range, $v\alt \sqrt{\lambda}M_{pl}$, in which large amplification of fluctuations is expected via tachyonic instability. 
Hence instead of doing simulations for such a small value of $ v/M_{pl} $, we adopt $ \lambda = 10^{-4} $ and 
$ v=10^{-2} M_{pl} $, so that both two scales are included in our simulations. 
We note that in the double inflation scenario \cite{Kofman:1985aw,Silk:1986vc}  $ \lambda > 10^{-12} $ can be realized.

We start our simulations at the time when $ \phi = v \frac{v}{M_{pl}} $. 
The initial condition for $ {\cal P}_{\delta \phi} $
is determined by solving the linearized equations,
\begin{equation}
{\ddot {\delta \phi}_k}+3 H {\dot {\delta \phi}_k}+\left( \frac{k^2}{a^2}+V''(\phi) \right) {\delta \phi}_k =0. \label{tachyonic3}
\end{equation}
from $ \phi \ll\phi_0 $ to $ \phi = \phi_0 $.
We mainly present the results for two-dimensional simulations 
($ 2048^2 $) with translation symmetry along the
$ z $-axis imposed to cover the still widely separated typical length scales of the different instabilities discussed above. Later on, we briefly comment on three dimensional cases.

\begin{figure}[thb]
\begin{center}
  \includegraphics[width=8cm,clip]{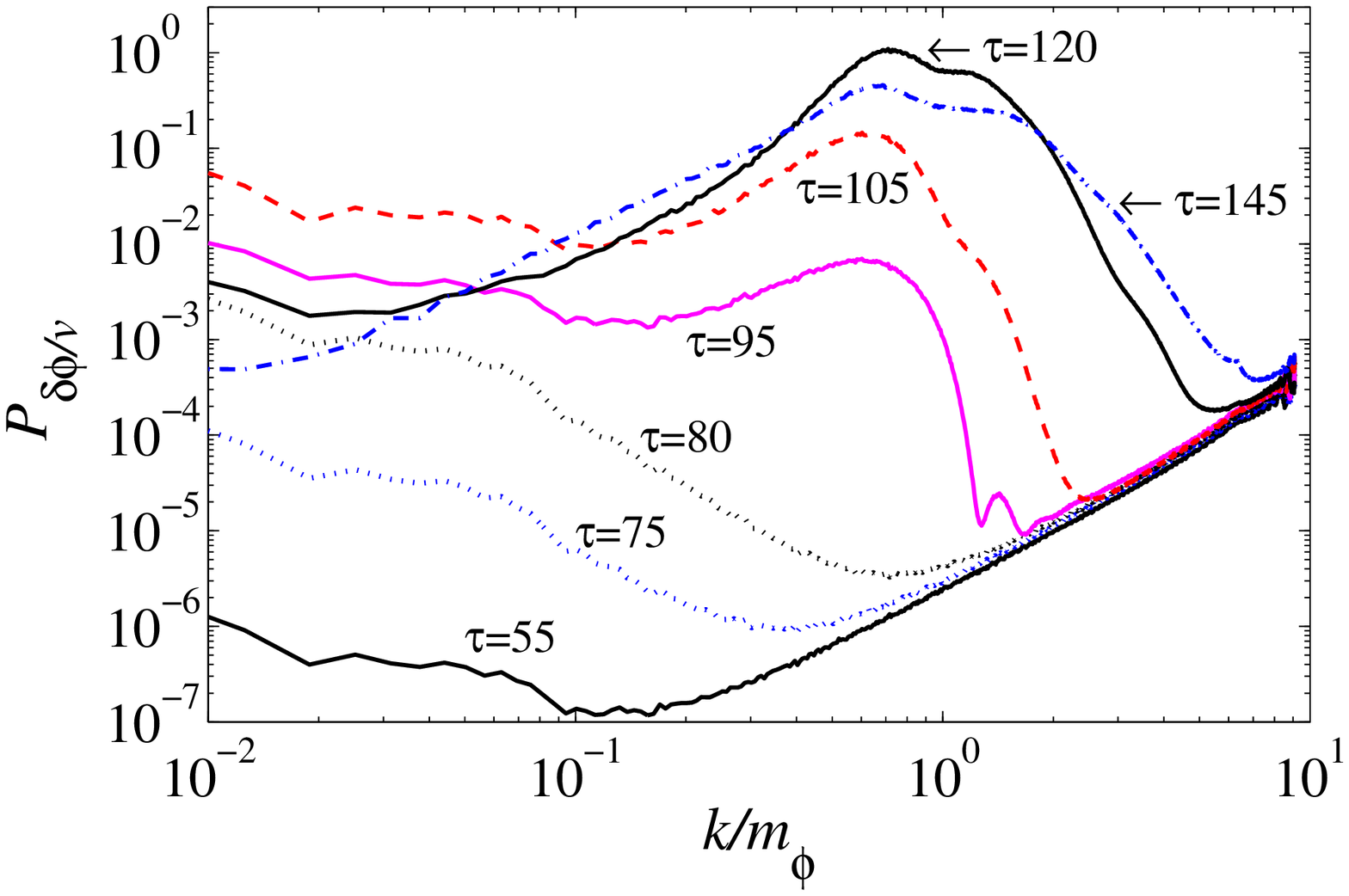}
\end{center}
\vspace*{-8mm}
\caption{Time evolution of $ {\cal P}_{\delta \phi /v} $ 
for the Coleman-Weinberg potential
for $ \lambda =10^{-4} $ and $ v/M_{pl}=10^{-2} $.
$ \tau $ is a dimensionless time defined by
$ \tau \equiv t/m_{\phi} $ where $ t $ is cosmological time.
The plot (also Fig.~2) is made by averaging ten different simulations in order
to reduce the statistical errors. 
}
\label{fig1} 
\begin{center}
  \includegraphics[width=8cm,clip]{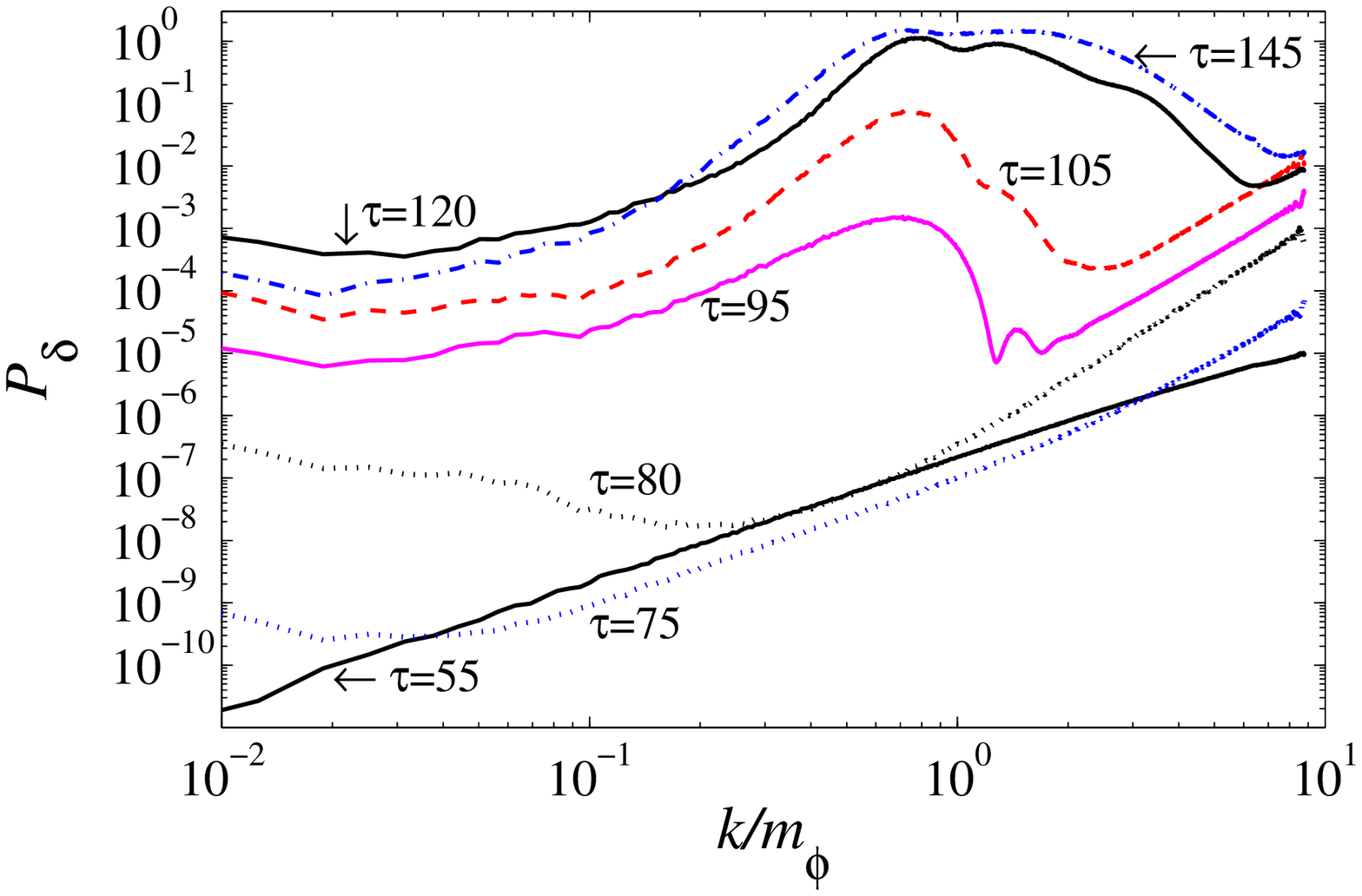}
\end{center}
\vspace*{-8mm}
\caption{Time evolution of $ {\cal P}_{\delta} $ for the Coleman-Weinberg potential
for $ \lambda =10^{-4} $ and $ v/M_{pl}=10^{-2} $. The 
horizon scale is located in the range $ -1.8 < {\log}_{10} k/m_{\phi} < -2.0 $
(slightly outside the range of our simulation) 
during preheating.
}
\label{fig2} 
\end{figure}

\section{Numerical results}
\subsection{$\lambda=10^{-4}, v/M_{pl}=10^{-2}$}
This is a marginal case in which non-linear interactions become efficient just when the inflaton reached the minimum of the potential.
But as we will see, simulations show that the linear regime lasts a little bit longer until the middle stage of the parametric resonance.
Fig.~\ref{fig1} and Fig.~\ref{fig2}, respectively, show the evolution of $ {\cal P}_{\delta \phi}/v $ and
$ {\cal P_{\delta}} $ for 
$ \lambda =10^{-4} $ and $ v/M_{pl}=10^{-2} $.
Here the proper time $\tau$ is measured in units of the inverse of the typical mass scale of the inflaton potential 
$m_\phi\equiv \sqrt{\lambda}v$. 
Let us first focus on Fig.~\ref{fig1}.
We find that long wavelength perturbations are selectively amplified during the tachyonic phase as expected from the linear analysis. 
After $ \phi $ starts to oscillate around the potential minimum 
{\bf ($\tau \gtrsim90$)}, parametric resonance occurs at scales around 
$ k/a \sim m_{\phi} $.
A remarkable thing is that once parametric resonance commences, the amplitude of $\delta\phi$ with long wavelengths decreases in time, and its spectrum eventually obeys the power law $ \propto k^2 $.
This scaling can be interpreted as a result of losing correlations on large scales by parametric resonance on small scales.

Next we focus on fig.~\ref{fig2}.
We find that $ {\cal P}_{\delta} $ around horizon scales ($k\sim aH$) becomes as large as $ 10^{-3} $, which is in good agreement with the prediction of the linear approximation.
In the linear approximation, we have 
$ \delta \sim \frac{\dot{\rho}}{\rho H} \delta N \sim \delta N$ on large scales, where $ \delta N $ is the perturbation of the $e$-folding number on flat hypersurfaces and is roughly estimated as $\sqrt{\lambda}$ at the initial time ($\phi=\phi_o$). 
Here we used the fact that $\delta N$ is conserved on large scales for single field models. 
Although the linear perturbation theory breaks down after the tachyonic instability, lattice simulations suggest that there is no further amplification of $\delta$ on sub-horizon scales by non-linear effects in this case.

\begin{figure}[tbh]
\begin{flushleft}
 \includegraphics[width=4.2cm,clip]{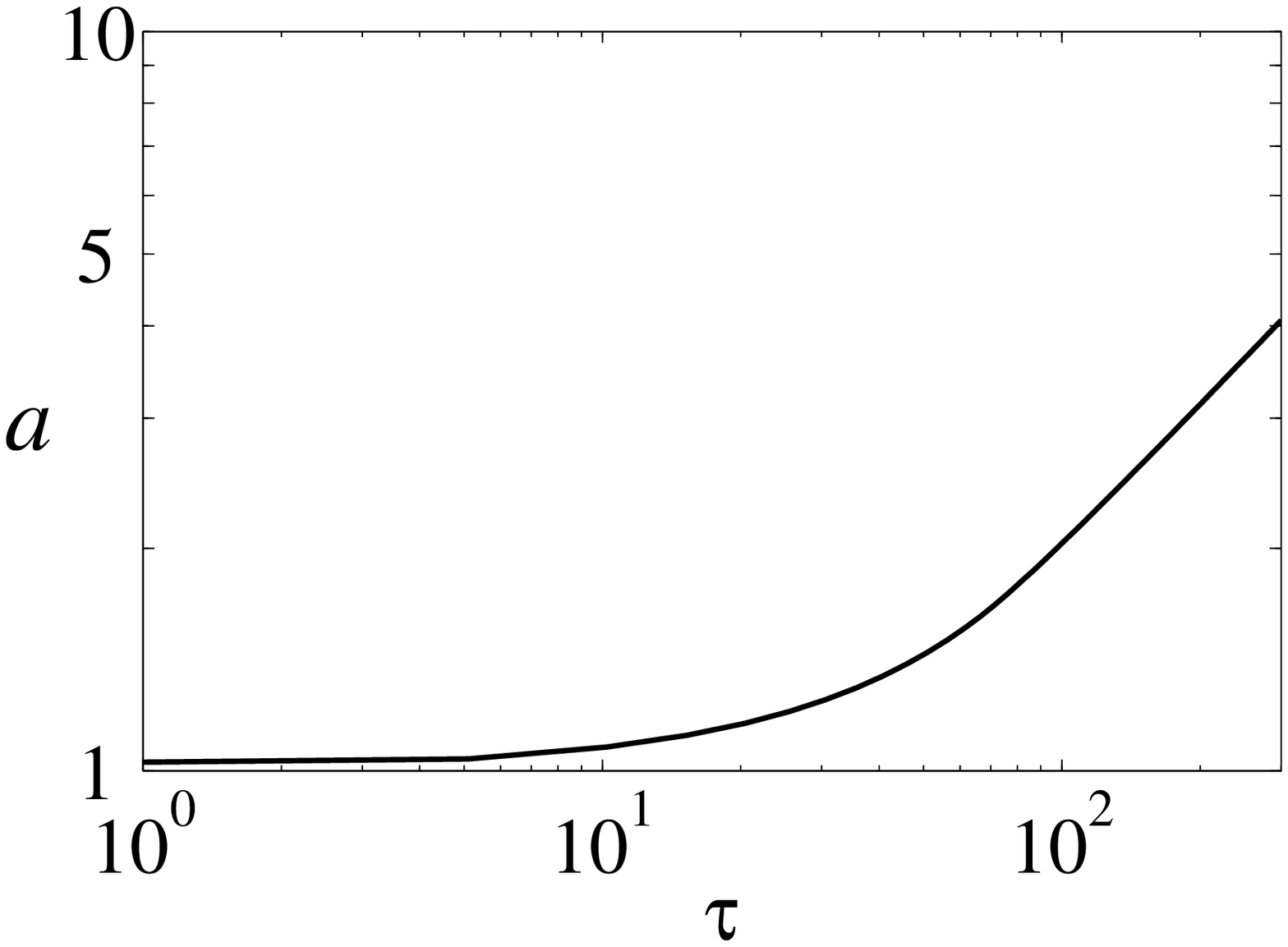}
 \includegraphics[width=4.2cm,clip]{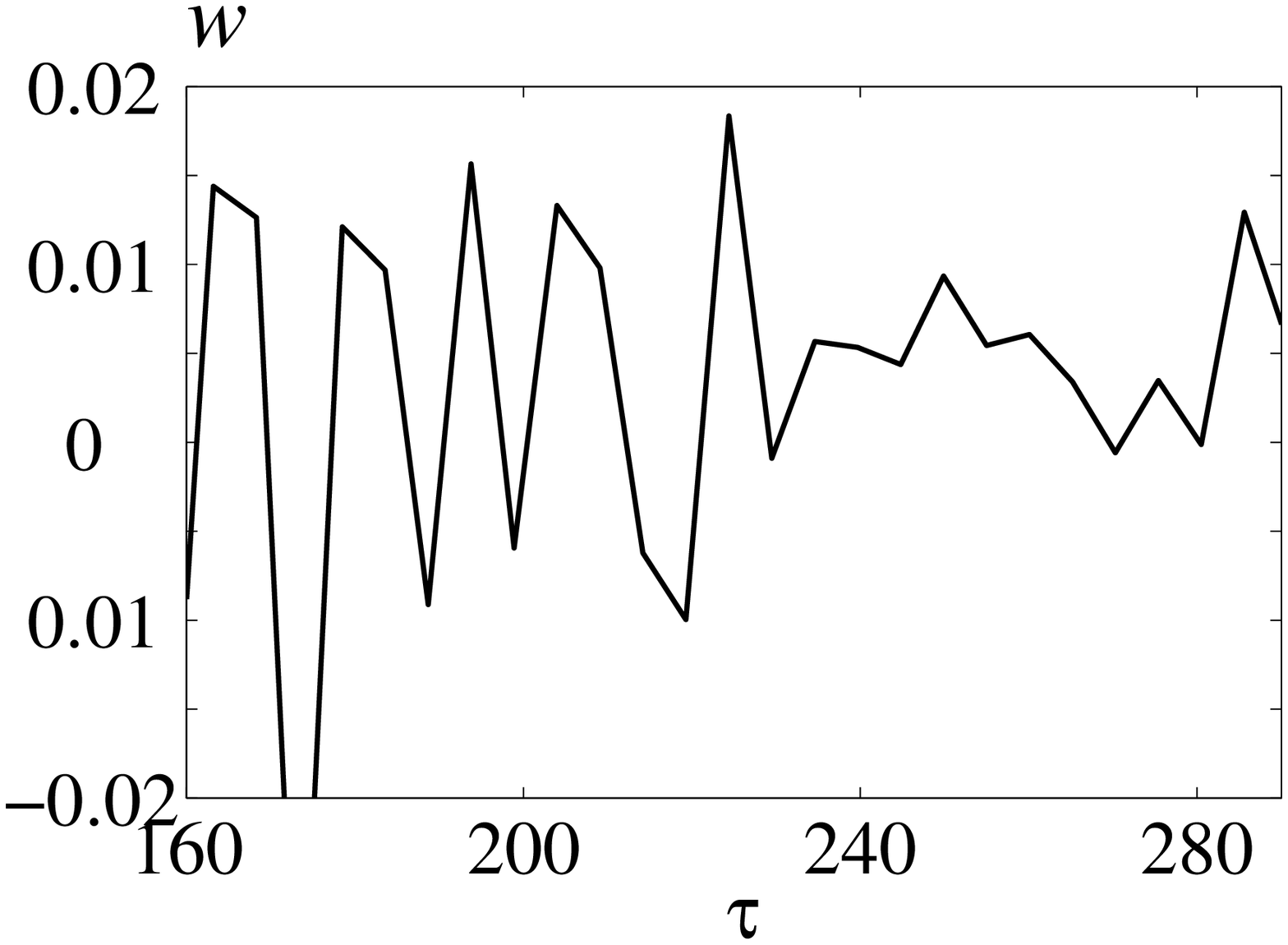}
\end{flushleft}
\vspace*{-8mm}
\caption{Evolution of the scale factor $a(\tau)$ and $w$.
}
\label{fig3} 
\end{figure}

Immediately after the parametric resonance starts, the universe begins to expand almost like the dust-dominated universe because the energy density of the universe is dominated by the long wavelength modes compared with the mass scale, which are selectively excited by the parametric resonance. 
Figure ~\ref{fig3} shows the evolution of scale factor $a$ and $ w $ respectively, where $ w = P/\rho $.
As shown in \cite{Carr:1975qj} by the extrapolation of the linear approximation, the lowest smoothed density contrast $ {\delta}_c $ above which BHs are formed is about $ w $.
Numerical simulations \cite{Shibata:1999zs,Musco:2004ak} which studied BH formation in the radiation dominated era showed $ 0.3 \lesssim {\delta}_c \lesssim 0.5$, supporting the conclusions of \cite{Carr:1975qj}.
In the following estimate of the produced BH abundance, we adopt this threshold so that we assume the relation $ {\delta}_c = w $. 
Assuming that the probability distribution of density perturbations at the Jeans wavelength $ \sim \sqrt{w} H^{-1} $ is Gaussian, the observational upper limit of the standard deviation of this distribution 
$ \sigma $ is about $ 0.1 {\delta}_c $ for a wide range of BH mass scales.
We find from Fig.~\ref{fig3} that the time average of $ w $ is smaller than $ 10^{-2} $. 
Therefore the threshold value on $ {\cal P}_{\delta} $ is  $\sim (0.1 w)^2 \approx 10^{-6} $, which is much smaller than the values we find from our numerical simulations; see fig.~\ref{fig2}. In this case, BHs are over-produced.

\begin{figure}[bth]
\begin{center}
  \includegraphics[width=6cm,clip]{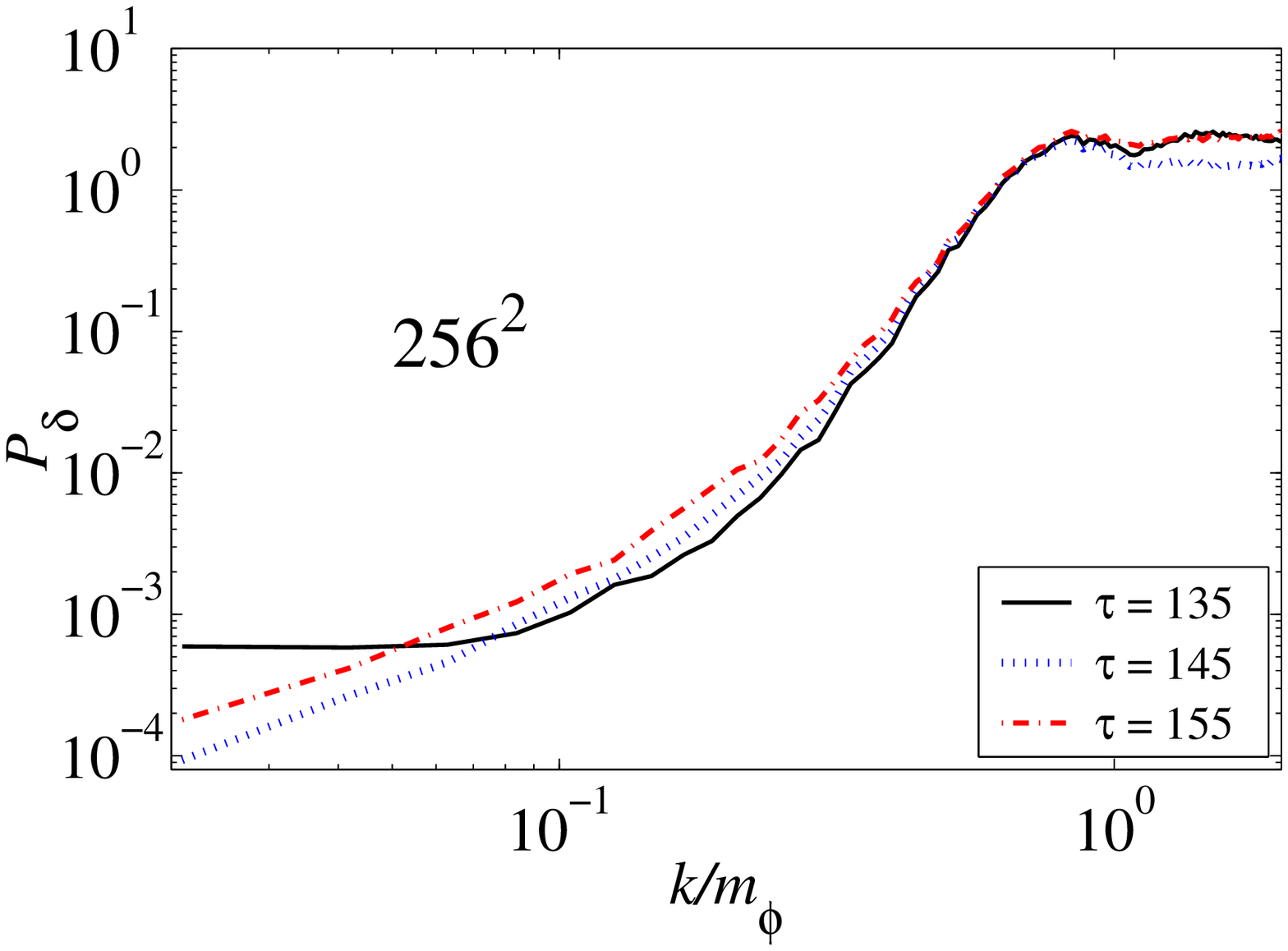}
  \includegraphics[width=6cm,clip]{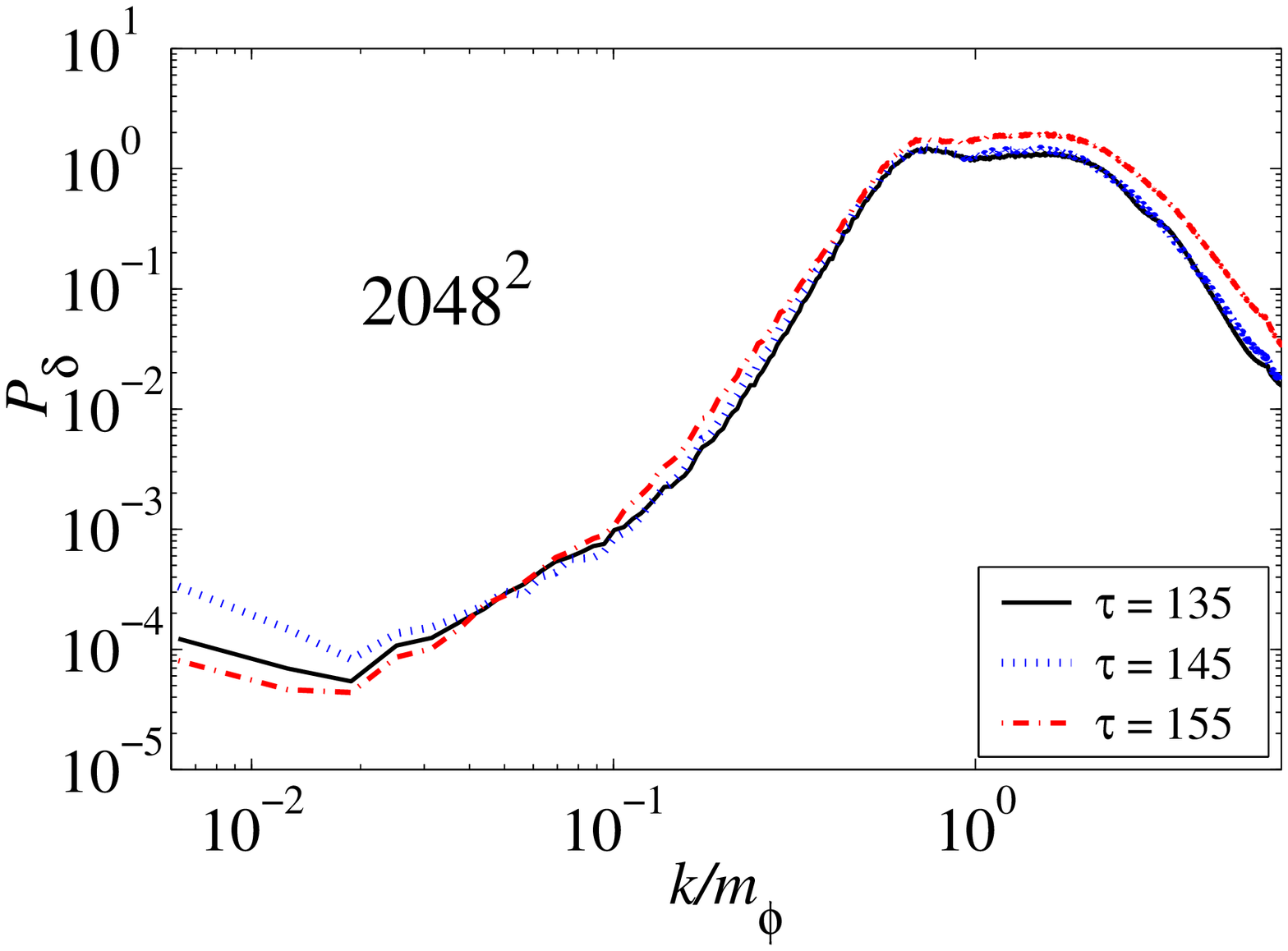}
\end{center}
\vspace*{-8mm}
\caption{Comparison of two-dimensional simulations with 
different grid sizes $(256^2)$ and $(2048^2)$. 
In both cases the time evolution of $ {\cal P}_{\delta} $ 
is plotted for $ \lambda =10^{-4} $ and $ v/M_{pl}=10^{-2} $.
}
\label{fig4} 
\end{figure}

We also performed simulations for different values of $ v/M_{pl} $ around $ 10^{-2} $ such as $ 5 \times 10^{-2}, 5 \times 10^{-3}, 2 \times 10^{-3}$ and so on. All the cases exhibit similar behavior as Figs.~\ref{fig1} and ~\ref{fig2}. 

During the tachyonic phase, $ \delta $ also starts to grow following the amplification  of $ \delta \phi $ at large scales, and the spectrum ${\cal P}_\delta$ becomes slightly red.
Interestingly, at the late phase of the parametric resonance ($\tau\agt 120$ in fig.~\ref{fig2}) fluctuations of the $\phi$-field on large scales decrease and tend to obey the scaling 
$ {\cal P}_{\delta \phi} \propto k^2 $, 
while density perturbations on large scales 
remain close to the flat spectrum.
After $ \tau > 145 $, $ {\cal P}_{\delta} $ becomes almost time-independent.
This will be explained by causality. 
Energy transfer faster than the velocity of
light cannot occur. If we neglect the presence of 
correlations between different Fourier modes of 
$ \delta \phi $, 
the red spectrum seen in $ {\cal P}_{\delta} $ on large scales 
at a late time after parametric resonance cannot be 
explained from the blue spectrum of $ {\cal P}_{\delta \phi/v} $.  
This means that there are strong correlations among
different Fourier modes. 
As is expected, if we increase the amplitude of 
initial fluctuations on large scales by hand, 
the resulting $ {\cal P}_{\delta} $ also 
becomes larger.

\begin{figure}[htb]
\begin{center}
  \includegraphics[width=7cm,clip]{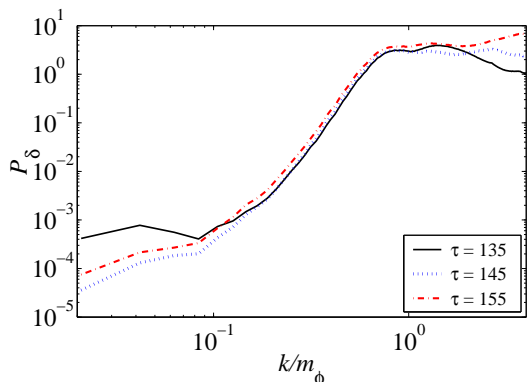}
\end{center}
\vspace*{-8mm}
\caption{
Three-dimensional time evolution of $ {\cal P}_{\delta} $ 
for $ \lambda =10^{-4} $ and $ v/M_{pl}=10^{-2} $.
}
\label{fig5} 
\end{figure}

We also performed three-dimensional simulations 
with grid size $ (256^3) $, which is not sufficient to include both the scales of tachyonic instability and of parametric resonance.
Before showing the result of three-dimensional simulations, we present comparison of two-dimensional simulations with different grid sizes $(256^2)$ and $(2048^2)$ in fig.\ref{fig4}. 
In the former case the scales at which parametric resonance is efficient, $ k/a \sim m_{\phi} $, are not fully covered by the simulation. 
Then, the amplitude on large scales is relatively enhanced
(but only by a small factor), presumably because the tachyonic preheating is kept efficient for a longer period due to weaker backreaction. 
Fig.~\ref{fig5} shows $ {\cal P}_{\delta} $ after the tachyonic instability for three-dimensional case. We see that this spectrum is almost identical to the two-dimensional case with $N=256$ presented in fig.~\ref{fig4}, and its growth has already been saturated on large scales. 
This suggests that three dimensional simulations with
$ 2048^3 $ should yield a spectrum similar to the corresponding two-dimensional one and thus there is a strong potential of over-producing BHs in the three-dimensional case. 
 
\subsection{$\lambda=1, v/M_{pl}=10^{-2}$}

\begin{figure}[htb]
\begin{center}
  \includegraphics[width=7cm,clip]{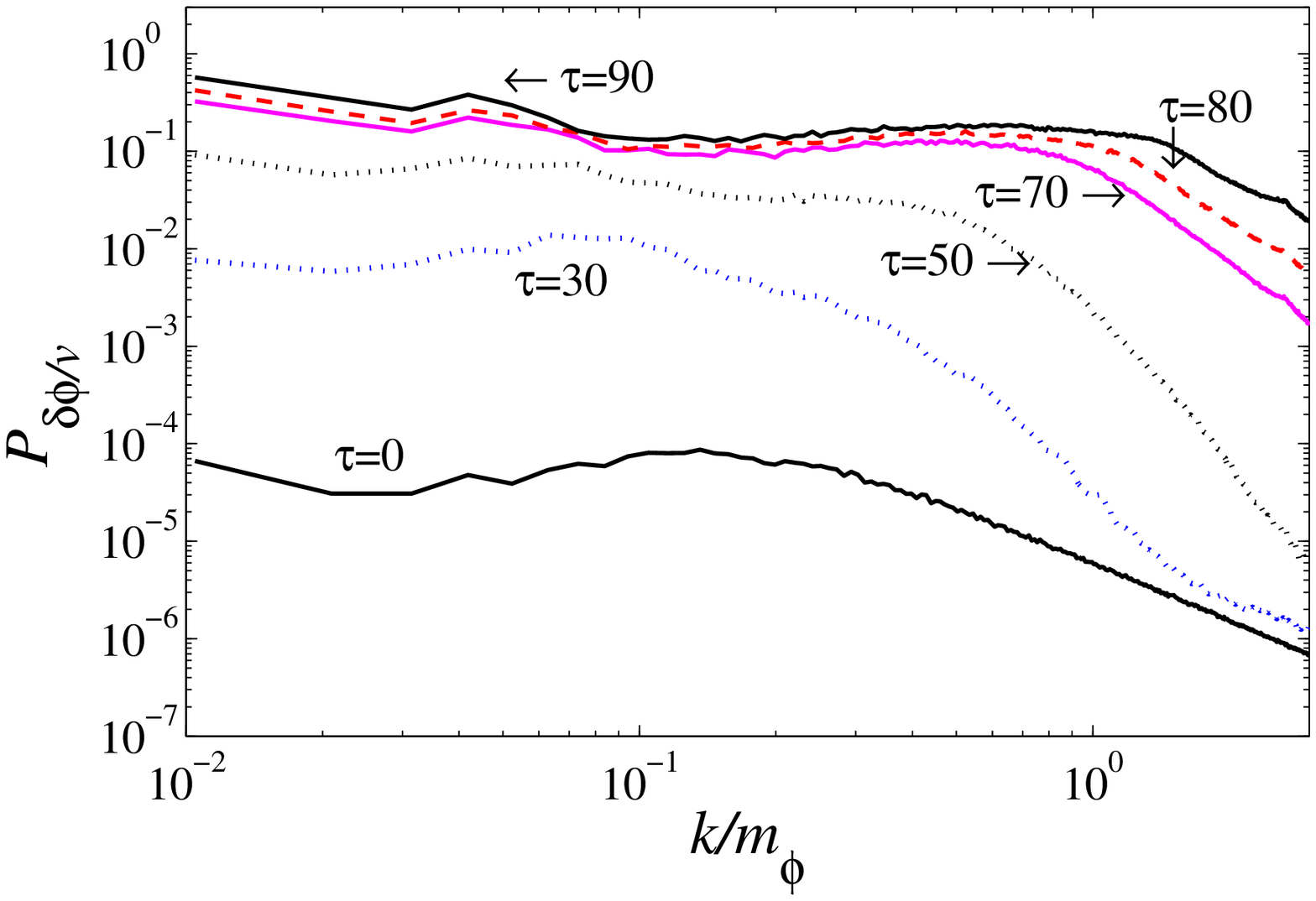}
\end{center}
\vspace*{-8mm}
\caption{
Time evolution of $ {\cal P}_{\delta \phi/v} $ 
for $ \lambda =1 $ and $ v/M_{pl}=10^{-2} $. 
}
\label{fig6} 
\begin{center}
  \includegraphics[width=7cm,clip]{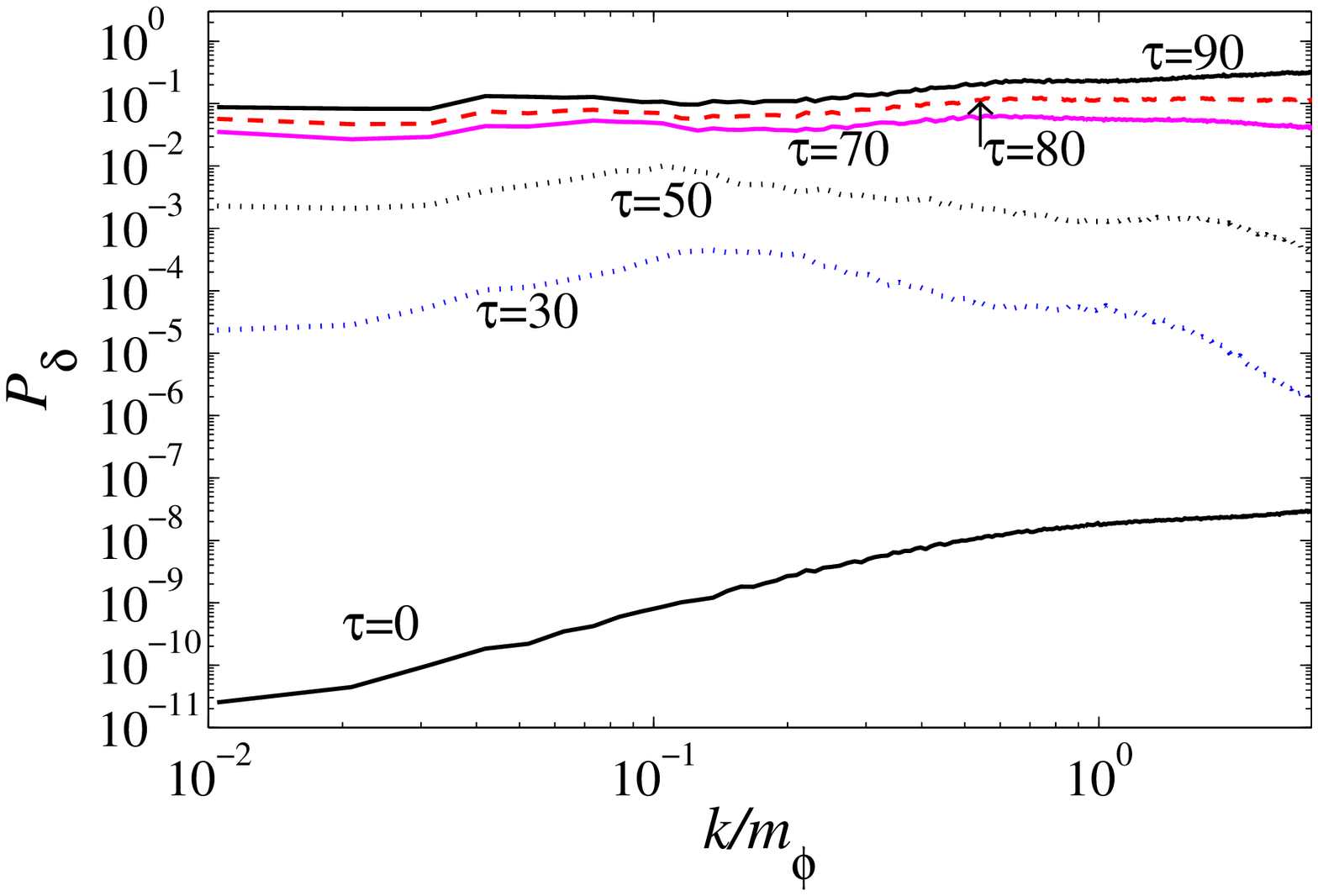}
\end{center}
\vspace*{-8mm}
\caption{
$ {\cal P}_{\delta} $ for $ \lambda =1 $ and $ v/M_{pl}=10^{-2} $.
}
\label{fig7} 
\end{figure}

Fig.~\ref{fig6} and \ref{fig7} show the evolution of
$ {\cal P}_{\delta \phi/v}$ and $ {\cal P}_{\delta} $ respectively for $\lambda=1, v/M_{pl}=10^{-2}$. 
In this case, non-linear interactions are already important during the tachyonic phase.
We have removed the contribution to $ \langle {\phi}^2 \rangle $ from the vacuum fluctuations of the short wavelength modes by introducing the cutoff, 
${\cal P}_{\delta \phi} \longrightarrow 
{\cal P}_{\delta \phi}/(1+{(k/k_c)}^{\alpha})$.
This cutoff is necessary because otherwise the dynamics is largely affected by the shortest wavelength modes which are not resolved well numerically.

From fig.~\ref{fig6},
we see that, contrary to the case with 
$ \lambda =10^{-4}$ and $ v/M_{pl}=10^{-2} $
where linear approximation is valid during the tachyonic phase and 
the subsequent parametric resonance occurs efficiently,
parametric resonance is less efficient in this case
because non-linear interactions among modes around
$k\lesssim m_{\phi}$ have become important when inflaton
reaches the potential minimum.
In particular,
$ {\cal P}_{\delta \phi/\phi} (k) $ at small $ k $ does not approach the scaling regime given by 
$ {\cal P}_{\delta \phi} \propto k^2 $, and long range correlations are still maintained after parametric resonance.

Next we focus on the evolution of the density perturbations.
We found that ${\cal P}_{\delta}$ at a small $k$ almost remains constant 
after $\tau \gtrsim 90$.
The value of ${\cal P}_{\delta}$ at small $k$ is about $0.1$ which is about one order of magnitude smaller than the one in the linear approximation.
This shows that the early importance of non-linear interactions  
suppress the terminal amplitude of density perturbations presumably
due to the effect of backreaction rather than increase it.

Before closing this section,
we briefly comment on the possibility of producing BHs for 
other parameter regions of $\lambda$ and $v/M_{pl}$ which
are not covered in the simulations.
The results of the numerical simulations of both A and B
show that the non-linear interactions which may become
important during the phase of the tachyonic instability or
that of the parametric resonance do not amplify the density
perturbations and hence the final curvature perturbations on 
constant energy density hypersurface do not exceed the one 
evaluated by the linear perturbation theory where the curvature 
perturbations is given by $\sim \sqrt{\lambda}$. 

If we assume that this result holds for smaller values of 
$\lambda$ and $v/M_{pl}$ which are not covered in the simulations,
then the amplitudes of perturbations evaluated by the linear 
perturbation theory, which is $\sqrt{\lambda}$ in this case, will 
determine the abundance of produced BHs.
If we also assume that the threshold value of over-production of
BHs does not depend on $\lambda$ and $v/M_{pl}$,
then BHs will indeed be over-produced for $\lambda \gtrsim 10^{-6}$.

\section{Discussions}
In summary, we studied non-linear evolution of density perturbations
on sub-horizon scales by two and three dimensional fully nonlinear lattice simulations.
We adopted the Coleman-Weinberg potential as a model of 
tachyonic instability, since 
this model is expected to capture typical features 
in more extended preheating models.

We found that the density perturbations below the horizon scale are 
selectively amplified by tachyonic instability
and thereafter maintain their amplitude, while fluctuations 
of the inflaton field decrease in time if the following parametric 
resonance dominates the perturbations. 
If the non-linear interactions become significant during
the tachyonic phase,
parametric resonance is less efficient and long range
correlations of inflaton fluctuations do not decay.

We also found that there is no enhancement of perturbations 
due to the non-linear interactions and the terminal value 
of the spectrum ${\cal P}_{\delta}$ on subhorizon scales 
is bounded from above by the linear estimate 
$\delta N^2\sim \lambda$.
Hence if we extrapolate this result to the realistic $\lambda \sim 10^{-10}$
determined by temperature fluctuations of the cosmic microwave
background,
we see that BH over-production is unlikely to occur for realistic $\lambda$ with $\delta N \sim 10^{-5}$.

We also found that the resulting power spectrum 
of density perturbations 
strongly depends on the initial
conditions when preheating is dominated by the
tachyonic instability.
The latter point is in contrast to 
the cases dominated by parametric 
resonance, in which the spectrum 
tends to obey a universal power law ${\cal P}_\delta \propto k^D $ 
($ D $ is the spatial dimension) independently 
of the initial shape of the spectrum.

Finally we briefly comment on the role of metric perturbations
which are not taken into account in our simulations.
It is necessary to include metric perturbations for the 
evolution of long wavelength perturbations.
In linear perturbation theory,
the evolution equations for inflaton fluctuations on flat
slicing is given by \cite{Sasaki:1995aw}
\begin{equation}
{\ddot {\delta \phi}}+3H {\dot {\delta \phi}} +\frac{k^2}{a^2}{\delta \phi}+V'' \delta \phi=a^{-3} {\left( \frac{a^3}{H} {\dot{\phi}}^2 \right)}^{\cdot} \delta \phi. \label{dis1}
\end{equation}
The right hand side of eq.~(\ref{dis1}) 
represents the contributions from 
metric perturbations.
We solved eq.~(\ref{dis1}) numerically and compared the
resulting power spectrum with the one without metric perturbations. 
We found little difference between the two power spectra,
which shows that the neglecting the metric perturbation is a
good approximation within linear perturbation theory.

There is a possibility that non-linearity may
enhance the perturbations when metric perturbations are 
taken into account but this possibility is beyond the scope of the current work.

T.S. thanks Misao Sasaki and Takashi Nakamura for useful
comments. This work is supported in part 
by Grant-in-Aid for Scientific Research, Nos.
14047212 and 16740141, 
and by that for the 21st Century COE
"Center for Diversity and Universality in Physics" at 
Kyoto university, both from the Ministry of
Education, Culture, Sports, Science and Technology of Japan.

\end{document}